\documentclass[12pt]{article}

\usepackage{graphicx}
\usepackage{amsmath,amssymb}
\usepackage{bm}
\usepackage{graphicx, color}
\usepackage{wrapfig}
\begin{document}
\title{
\begin{flushright}
\ \\*[-80pt] 
\begin{minipage}{0.2\linewidth}
\normalsize
KUNS-2166 \\*[50pt]
\end{minipage}
\end{flushright}
{\Large \bf 
Lepton Flavor Model from  $\Delta(54)$  Symmetry
\\*[20pt]}}

\author{
\centerline{
Hajime~Ishimori$^{1,}$\footnote{E-mail address: 
ishimori@muse.sc.niigata-u.ac.jp}, 
  \quad Tatsuo~Kobayashi$^{2,}$\footnote{E-mail address: 
kobayash@gauge.scphys.kyoto-u.ac.jp},
\quad Hiroshi Okada$^{3,}$
\footnote{E-mail address: HOkada@Bue.edu.eg},
}\\
 \centerline{Yusuke~Shimizu$^{1,}$\footnote{E-mail address: 
shimizu@muse.sc.niigata-u.ac.jp},
\quad and \quad  Morimitsu~Tanimoto$^{4,}$\footnote{E-mail address: 
tanimoto@muse.sc.niigata-u.ac.jp} }
\\*[20pt]
\centerline{
\begin{minipage}{\linewidth}
\begin{center}
$^1${\it \normalsize
Graduate~School~of~Science~and~Technology,~Niigata~University, \\ 
Niigata~950-2181,~Japan } \\
$^2${\it \normalsize 
Department of Physics, Kyoto University, 
Kyoto 606-8502, Japan} \\
$^3${\it \normalsize 
Centre for Theoretical Physics, The British University in Egypt, 
El-Sherouk City, 11837,  Egypt} \\
$^4${\it \normalsize
Department of Physics, Niigata University,~Niigata,  950-2181, Japan } 
\end{center}
\end{minipage}}
\\*[50pt]}
\vskip 2 cm
\date{\small
\centerline{ \bf Abstract}
\begin{minipage}{0.9\linewidth}
\medskip 
We present  the lepton flavor model with $\Delta (54)$,
 which  appears typically in heterotic string models 
on  the $T^2/Z_3$ orbifold. 
Our model reproduces the tri-bimaximal mixing in 
the parameter region around 
degenerate neutrino masses or two massless neutrinos.
We predict the deviation from the tri-bimaximal mixing
by putting the experimental data of neutrino masses
 in the normal hierarchy of neutrino masses.
  The upper bound of $\sin^2\theta_{13}$ is $0.01$.
   There is the strong correlation between  
 $\theta_{23}$ and $\theta_{13}$.
Unless $\theta_{23}$ is deviated from the maximal mixing considerably,
$\theta_{13}$ remains  to be tiny.
\end{minipage}
}

\begin{titlepage}
\maketitle
\thispagestyle{empty}
\end{titlepage}

\section{Introduction}
It is the important task to find  an origin of the observed hierarchies 
 in masses and flavor  mixing  for quarks and leptons.
Neutrino experimental  data provide us a valuable  clue to find this origin. 
In particular, recent experiments of the neutrino
 oscillation go into the new  phase  of precise  determination of
 mixing angles and mass squared  differences \cite{Threeflavors}. 
Those indicate the  tri-bimaximal mixing  for three flavors 
 in the lepton sector \cite{HPS}. 
Therefore, it is necessary
to find a natural model that leads to this mixing pattern with good accuracy.

The tri-bimaximal mixing  for three flavors indicates the specific
neutrino  mass matrix, in which matrix elements are connected each other.
 The  non-Abelian discrete flavor symmetry is  appropriate  to explain
 such a structure of the mass matrix  which leads to the tri-bimaximal,
because  the  symmetry provides the definite meaning of generations
 and connects different generations.
Actually, several types of models with various non-Abelian discrete flavor 
symmetries have been proposed, such as 
$S_3$ \cite{S3-FTY}-\cite{S3-Lavoura}, 
 $D_4$ \cite{Grimus}-\cite{Ishimori:2008ns}, $D_6$ \cite{D6},
$Q_4 \cite{Q4},\ Q_6$ \cite{Q6},
$A_4$ \cite{A4-Ma and Rajasekaran}-\cite{A4-Lin}, $T^\prime $ \cite{T-Feruglio}-\cite{T-Ding}, 
$S_4$ \cite{S4-Ma}-\cite{S4-Morisi} and 
$\Delta (27)$ \cite{Delta(27)-Gerard}-\cite{Delta(27)-Ma(2008)}.

Non-Abelian discrete symmetries are symmetries of 
geometrical solids.
Thus, an origin of non-Abelian discrete flavor symmetries 
may be compact extra dimensions, e.g. 
string-derived compact spaces.
Recently, which types of non-Abelian discrete flavor symmetries 
can appear in heterotic orbifold models has been studied 
\cite{Kobayashi:2004ya,Kobayashi:2006wq,Ko:2007dz}.
Simple orbifolds can lead to non-Abelian flavor symmetries such as 
$D_4$, $\Delta(54)$ and so on.
For example, the  $\Delta(54)$ flavor symmetry can appear 
typically in heterotic string models on 
factorizable orbifolds including the $T^2/Z_3$ orbifold. 
Other string compactifications would lead to different 
flavor symmetries.

The $D_4$ flavor model  has been already proposed  by
Grimus and Lavoura  \cite{Grimus} and phenomenologically important results 
have been obtained \cite{D4-2}.
The $\Delta(54)$ flavor symmetry would be also interesting, 
e.g. from the viewpoint that $\Delta(54)$ has 
triplet irreducible representations \cite{delta54}, while 
$D_4$ has only singlets and doublets.
Indeed, non-Abelian flavor symmetries,   
$A_4$, $S_4$, and $T'$, include triplet irreducible representations 
and those are useful to explain the  three generations of 
leptons with their mixing angles and reproduce 
the  tri-bimaximal mixing of flavors.
The $\Delta(54)$ flavor symmetry would have similarly 
interesting aspects.
However, the group $\Delta(54)$ is rather unfamiliar compared to other
discrete groups used as the flavor symmetry. 
Its phenomenological applications have not been studied.
Thus, our purpose in this paper is to present
 a lepton flavor model with the $\Delta(54)$ symmetry
and  study phenomenological implications.

The paper is organized as follows:
we present the framework of the lepton flavor model with $\Delta (54)$ 
in section 2, and 
 discuss the effect of the higher order corrections, in section 3.
In section 4, we present the potential analysis to assure the VEVs used 
in section 2. 
Numerical results are exhibited in section 5 for neutrino masses and mixing
 angles.
Section 6 is devoted to summary and discussion.
In the appendix, we present the character table, the kronecker products and 
 Clebsch Gordan coefficients of $\Delta (54)$.

\section{$\Delta(54)$ Lepton Flavor model}
In this section, we present the lepton flavor model with 
the $\Delta(54)$ flavor symmetry.
We propose our model within the framework of supersymmetric models.
However, similar non-supersymmetric models could be constructed.

The $\Delta(54)$ group is one of series of $\Delta(6n^2)$
that has been discussed by a few authors \cite{delta54,Lam}. The group
$\Delta(54)$ has irreducible representations $1_1$, $1_2$, $2_1$, $2_2$,
$2_3$, $2_4$, $3_1^{(1)}$, $3_1^{(2)}$, $3_2^{(1)}$,
and $3_2^{(2)}$.
It is remarked that there are four triplets and only $3_1^{(1)}\times 3_1^{(2)}$ leads to the trivial singlet. 
The relevant  multiplication rules are summarized in appendix.

\begin{table}[h]
\begin{center}
\begin{tabular}{|c|ccc||c||ccc|}
\hline
              &$(l_e,l_\mu,l_\tau)$ & $(e^c,\mu^c,\tau^c)$ & 
$(N_e^c,N_\mu^c,N_\tau^c)$         &$h_{u(d)}$ &$ \chi_1 $&  $(\chi_{2},\chi_3)$&  
$(\chi_{4},\chi_5,\chi_6)$ \\ \hline
$\Delta(54)$      &$3_1^{(1)}$      & $3_2^{(2)}$   & $3_1^{(2)}$          
& $1_1$ & $1_2$  & $2_1$ & $3_1^{(2)}$    \\
\hline
\end{tabular}
\end{center}
\caption{Assignments of $\Delta(54)$ representations}
\end{table}
Let us present the model of the lepton flavor with the $\Delta(54)$
group. The triplet representations of the group
correspond to the three generations of leptons.
The left-handed leptons $(l_e,l_\mu,l_\tau)$,  
the right-handed charged leptons  $(e^c,\mu^c,\tau^c)$
and the right-handed neutrinos $(N_e^c,N_\mu^c,N_\tau^c)$ 
are assigned by $3_1^{(1)}$,
$3_2^{(2)}$, and $3_1^{(2)}$, respectively. 
Since $3_1^{(1)}\times 3_1^{(2)}$ makes
trivial singlet $1_1$, only Dirac neutrino Yukawa couplings are allowed
in tree level. On the other hand, charged leptons and 
the right-handed Majorana neutrinos cannot have mass terms 
unless new scalars $\chi_i$ are introduced  in addition to the usual Higgs
doublets, $h_u$ and $h_d$. 
These new scalars are supposed to be $SU(2)$ gauge singlets.
The gauge singlets $\chi_1$, $(\chi_2, \chi_3)$ and
$(\chi_4, \chi_5, \chi_6)$ are assigned to
$1_2$, $2_1$, and $3_1^{(2)}$ of the $\Delta(54)$ representations,
respectively.
The particle assignments of $\Delta(54)$ are summarized in Table 1.
The usual Higgs doublets $h_u$ and $h_d$ are assigned to  
the trivial singlet $1_1$ of $\Delta(54)$.
Here, we use the conventional notation that 
we denote the superfield and its lowest scalar component by 
the same letter.

In this setup of the particle assignment,
let us consider the superpotential of leptons at the leading order
in terms of the cut-off scale $\Lambda$, which is taken to be
the Planck scale.
For charged leptons, the superpotential of 
the Yukawa sector respecting to $\Delta(54)$ symmetry
is given as 
\begin{eqnarray}
w_l
&=&
y_1^l 
 (  e^c l_e+ \mu^c l_\mu+  \tau^c l_\tau )\chi_1 h_d/\Lambda 
\nonumber\\
&&+y_2^l \ [  
 (\omega  e^c l_e+\omega^2  \mu^c l_\mu+  \tau^c l_\tau )\chi_2 
-( e^c l_e+\omega^2  \mu^c l_\mu+\omega \tau^c l_\tau )\chi_3]
\ h_d/\Lambda. 
\end{eqnarray}
For the right-handed Majorana neutrinos we can write the superpotential
as follows:
\begin{eqnarray}
w_N
&=&y_1 ( N_e^c N_e^c\chi_4+ N_\mu^c N_\mu^c\chi_5+ N_\tau^c N_\tau^c\chi_6)
\nonumber\\&&
+y_2 \ [( N_\mu^c N_\tau^c+ N_\tau^c N_\mu^c)\chi_4
+( N_e^c N_\tau^c+ N_\tau^c N_e^c)\chi_5
+( N_e^c N_\mu^c+ N_\mu^c N_e^c)\chi_6].
\end{eqnarray}
The superpotential for the Dirac neutrinos has tree level contributions 
as
\begin{eqnarray}
w_D
&=&
y_D 
 \ (  N^c_e l_e+N^c_ \mu l_\mu+  N^c_\tau l_\tau )h_u \ .
\end{eqnarray}

We assume that the scalar fields, $h_{u,d}$ and $\chi_i$, develop 
their vacuum expectation values (VEVs) as follows:
\begin{eqnarray}
\left<h_u\right>=v_u, \   \left<h_d\right>=v_d,
\quad
\left<\chi_1\right>=u_1,
\  
\left<(\chi_2,\chi_3)\right>=(u_2,u_3),
\ 
\left<(\chi_4,\chi_5,\chi_6)\right>=(u_4,u_5,u_6).
\end{eqnarray}
Then, we obtain the diagonal mass matrix for   charged leptons
\begin{eqnarray}
M_l
 = 
y_1^lv_d 
\begin{pmatrix}\alpha_1  & 0 & 0 \\ 
           0  & \alpha_1  &  0  \\
                 0  & 0 & \alpha_1   \\
 \end{pmatrix} 
+y_2^l  v_d
\begin{pmatrix} \omega\alpha_2-\alpha_3 & 0 & 0 \\ 
               0    & \omega^2\alpha_2-\omega^2\alpha_3 &   0 \\
                 0 & 0 &  \alpha_2-\omega\alpha_3 \\
 \end{pmatrix}, 
\label{ME}
 \end{eqnarray}
while the right-handed Majorana mass matrix is given as 
\begin{eqnarray}
M_N
&=&
 {y_1  \Lambda}
\begin{pmatrix}\alpha_4  & 0 & 0 \\ 
               0    & \alpha_5  &0    \\
                 0  & 0 &\alpha_6   \\
 \end{pmatrix} 
 + {y_2 \Lambda}
\begin{pmatrix}0   & \alpha_{6} & \alpha_{5} \\ 
                   \alpha_{6}  & 0   &\alpha_{4}    \\
                   \alpha_{5}  & \alpha_{4}  & 0    \\
 \end{pmatrix},
\label{MR}
\end{eqnarray}
and the Dirac mass matrix of neutrinos is
\begin{eqnarray}
M_D = 
y_Dv_u
\begin{pmatrix} 1  & 0 & 0 \\ 
           0  &  1  &  0  \\
                 0  & 0 &  1   \\
 \end{pmatrix},
\label{MD}
 \end{eqnarray}
where we denote $\alpha_i=u_i/\Lambda \ (i=1-6)$.
By using the seesaw mechanism $M_\nu = M_D^T M_N^{-1} M_D$, the neutrino
mass matrix can be written as
\begin{eqnarray}
M_\nu
&=&
 \frac{y_D^2v_u^2}{ \Lambda d}
\begin{pmatrix}y_1^2\alpha_5\alpha_6-y_2^2\alpha_4^2  & -y_1y_2 
\alpha_6^2+y_2^2\alpha_4\alpha_5  & -y_1y_2 \alpha_5^2+y_2^2\alpha_4\alpha_6 \\ 
               -y_1y_2 \alpha_6^2+y_2^2\alpha_4\alpha_5   & 
y_1^2\alpha_4\alpha_6-y_2^2\alpha_5^2 & -y_1y_2 \alpha_4^2+y_2^2 \alpha_5\alpha_6    \\
              -y_1y_2 \alpha_5^2+y_2^2\alpha_4\alpha_6 & -y_1y_2 
\alpha_4^2+y_2^2 \alpha_5\alpha_6 &  y_1^2\alpha_4\alpha_5-y_2^2\alpha_6^2   \\
 \end{pmatrix}, 
\label{neumassmatrix}
 \nonumber\\
 \nonumber\\
 d&=&y_1^3\alpha_4\alpha_5\alpha_6-y_1y_2^2\alpha_4^3-
y_1y_2^2\alpha_5^3-y_1y_2^2\alpha_6^3
+2y_2^3\alpha_4\alpha_5\alpha_6.
\end{eqnarray}

   Since the charged leptons  mass matrix is diagonal one,
we can simply get the mass eigenvalues as
\begin{eqnarray}
\left(  \begin{array}{cc}
m_e  \\ 
m_\mu  \\ 
m_\tau  \\ 
\end{array} \right)
=
 v_d
\left(  \begin{array}{ccc}
1&\omega &-1  \\ 
1&\omega^2 &-\omega^2  \\ 
1&1 &-\omega  \\ 
\end{array} \right)
\left(  \begin{array}{cc}
y^\ell_1\alpha_1  \\ 
y^\ell_2\alpha _2  \\ 
y^\ell_2\alpha _3  \\ 
\end{array} \right) .
\end{eqnarray}
In order to estimate  magnitudes of $\alpha_1$,  $\alpha_2$ and  $\alpha_3$,
we  rewrite  as
\begin{eqnarray}
\left(  \begin{array}{cc}
y^\ell_1\alpha_1  \\ 
y^\ell_2\alpha _2  \\ 
y^\ell_2\alpha _3  \\ 
\end{array} \right)
=
\frac{1 }{3v_d}
\left(  \begin{array}{ccc}
1&1 & 1  \\ 
-\omega-1&\omega & 1   \\ 
-1&-\omega &\omega+1  \\ 
\end{array} \right)
\left(  \begin{array}{cc}
m_e \\ 
m_\mu  \\ 
m_\tau  \\ 
\end{array} \right) ,
\end{eqnarray}
which gives the relation of $|y^\ell_2\alpha_2| = |y^\ell_2\alpha_3|$.
Inserting the experimental values of the charged lepton masses
and $v_d\simeq 55$GeV, which is given by taking $\tan\beta=3$,
we obtain numerical results
\begin{equation}
\begin{pmatrix}
y_1^\ell \alpha _1 \\
y_2^\ell \alpha _2 \\
y_2^\ell \alpha _3 
\end{pmatrix} = \begin{pmatrix}
                         1.14\times 10^{-2} \\                                   
                         1.05\times 10^{-2} e^{0.016 i\pi } \\      
                         1.05\times 10^{-2} e^{0.32 i\pi }   \         
\end{pmatrix}.
\label{alpha123}
\end{equation}
Thus, it is found that  $\alpha_i(i=1,2,3)$ are order of ${\cal O}(10^{-2})$
 if the Yukawa couplings are order one.

In our model,
the lepton mixing comes from the structure of the neutrino mass matrix
of Eq.(\ref{neumassmatrix}).
 In order to reproduce the maximal mixing between
 $\nu_\mu$ and $\nu_\tau$,  we take $\alpha_5=\alpha_6$, and then 
we have
\begin{eqnarray}
\label{mass}
M_\nu 
 &=&
 \frac{y_D^2v_u^2}{ \Lambda d}
\begin{pmatrix}y_1^2\alpha_5^2-y_2^2\alpha_4^2  & -y_1y_2 
\alpha_5^2+y_2^2\alpha_4\alpha_5  & -y_1y_2 \alpha_5^2+y_2^2\alpha_4\alpha_5 \\ 
               -y_1y_2 \alpha_5^2+y_2^2\alpha_4\alpha_5   & 
y_1^2\alpha_4\alpha_5-y_2^2\alpha_5^2 & -y_1y_2 \alpha_4^2+y_2^2 \alpha_5^2    \\
               -y_1y_2 \alpha_5^2+y_2^2\alpha_4\alpha_5  & -y_1y_2 
\alpha_4^2+y_2^2 \alpha_5^2 &  y_1^2\alpha_4\alpha_5-y_2^2\alpha_5^2   \\
 \end{pmatrix}.
\label{neutri}
\end{eqnarray}
 
The tri-bimaximal mixing is realized by the condition of
$M_\nu(1,1)+M_\nu(1,2)=M_\nu(2,2)+M_\nu(2,3)$ in Eq. (\ref{neutri}), 
which turns to
\begin{eqnarray}
(y_1-y_2)(\alpha_4-\alpha_5)(y_1\alpha_5-y_2\alpha_4)=0.
\end{eqnarray}
Therefore, we have  three cases realizing the tri-bimaximal mixing
in Eq.(\ref{neutri}) as
\begin{eqnarray}
y_1=y_2, \qquad \alpha_4=\alpha_5, \qquad y_1\alpha_5=y_2\alpha_4.
\label{limit}
\label{condition}
\end{eqnarray}
Let us investigate the neutrino mass spectrum in these cases.
In general the neutrino mass matrix with the tri-bimaximal mixing is
expressed as
\begin{eqnarray}
M_\nu
=
\frac{m_1+m_3}{2}
 \begin{pmatrix}1   & 0 & 0 \\ 
                   0    & 1   &0    \\
                   0  & 0  & 1    \\
 \end{pmatrix}
+\frac{m_2-m_1}{3}
 \begin{pmatrix}1   & 1 & 1 \\ 
                   1    & 1   &1    \\
                   1  & 1  & 1    \\
 \end{pmatrix}
+\frac{m_1-m_3}{2}
 \begin{pmatrix}1   & 0 & 0 \\ 
                   0    & 0   &1    \\
                   0  & 1  & 0    \\
 \end{pmatrix} 
 .
\label{tri-bi1}
\end{eqnarray}
Actually, the neutrino mass matrix of
Eq.(\ref{neutri}) is decomposed under the condition in Eq.(\ref{condition})
as follows.
In the case of $\alpha_4=\alpha_5$, the neutrino mass matrix is expressed as
\begin{eqnarray}
M_\nu 
 &=&
 \frac{y_D^2v_u ^2\alpha_4^2(y_1-y_2)}{ \Lambda d}
\left[
(y_1+2y_2)
 \begin{pmatrix}1   & 0 & 0 \\ 
                   0    & 1   &0    \\
                   0  & 0 & 1    \\
 \end{pmatrix}
-y_2
 \begin{pmatrix}1   & 1 & 1 \\ 
                   1    & 1   &1    \\
                   1  & 1  & 1    \\
 \end{pmatrix}
\right ].
\end{eqnarray}
Therefore, it is found that neutrino masses are given as 
\begin{eqnarray}
\frac{m_1+m_3}{2}
&=&
 \frac{y_D^2v_u^2\alpha_4^2(y_1-y_2)}{ \Lambda d}
(y_1+2y_2),
\nonumber\\
\frac{m_2-m_1}{3}
&=&-\frac{y_D^2v_u^2 \alpha_4^2}{ \Lambda d}
(y_1-y_2) y_2,
\nonumber\\
m_1-m_3&=&0.
\end{eqnarray}
In the case of $y_1=y_2$, the mass matrix is decomposed as
\begin{eqnarray}
M_\nu 
 &=&
 \frac{y_D^2y_1^2v^2(\alpha_4-\alpha_5)}{ \Lambda d}
\left[
\alpha_5
 \begin{pmatrix}1   & 1 & 1 \\ 
                   1    & 1   &1    \\
                   1  & 1  & 1    \\
 \end{pmatrix}
-(\alpha_4+2\alpha_5)
 \begin{pmatrix}1   & 0 & 0 \\ 
                   0    & 0   &1    \\
                   0  & 1 & 0    \\
 \end{pmatrix}\right ], 
\end{eqnarray}
 and we have
\begin{eqnarray}
m_1+m_3&=&0,
\nonumber\\
\frac{m_2-m_1}{3}
&=&\frac{y_D^2 y_1^2 v_u^2(\alpha_4-\alpha_5)}{ \Lambda d}
\alpha_5,
\nonumber\\
\frac{m_1-m_3}{2}
&=&-\frac{y_D^2 y_1^2 v_u^2(\alpha_4-\alpha_5)}{ \Lambda d}
(\alpha_4+2\alpha_5).
\end{eqnarray}
In the last case of
$y_1\alpha_5=y_2\alpha_4$, we have
\begin{eqnarray}
M_\nu 
 &=&
 \frac{y_D^2v_u^2}{ \Lambda d}
y_1^2 \alpha_4\alpha_5 \left (1-\frac{\alpha_5^3}{\alpha_4^3}\right)
\left[
 \begin{pmatrix}1   & 0 & 0 \\ 
                   0    & 1   &0    \\
                   0  & 0 & 1    \\
 \end{pmatrix}
-
 \begin{pmatrix}1   & 0 & 0 \\ 
                   0   & 0   &1    \\
                   0  & 1  & 0    \\
 \end{pmatrix}
\right ]. 
\end{eqnarray}
Then, we obtain
\begin{eqnarray}
&&m_3=
 \frac{2y_D^2v_u^2}{ \Lambda d}
y_1^2 \alpha_4\alpha_5 \left (1-\frac{\alpha_5^3}{\alpha_4^3}\right),
\nonumber\\
&& m_2=m_1=0 \ .
\label{tri-bi2}
\end{eqnarray}

Thus, the tri-bimaximal mixing is not realized for
arbitrary neutrino masses $m_1$, $m_2$ and $m_3$ in our model.
In both conditions of $y_1=y_2$ and $\alpha_4=\alpha_5$,
we have $|m_1|=|m_3|$, which leads to quasi-degenerate neutrino masses
due to the condition of $\Delta m^2_{\rm atm} \gg \Delta m_{\rm sol}^2$.
Therefore, we do not discuss these cases in this paper
because we need fine-tuning of parameters
in order to be consistent with the experimental data of the neutrino
oscillations \cite{Threeflavors}.

In the case of $y_1\alpha_5=y_2\alpha_4$, the neutrino mass matrix turns to be
\begin{eqnarray}
M_\nu
=
 \frac{y_D^2y_1^2v_u^2}{ \Lambda d}
\begin{pmatrix}0  & 0  & 0 \\ 
               0   &  \alpha_4\alpha_5- \alpha_5^4/\alpha_4^2 & 
-\alpha_4\alpha_5+ \alpha_5^4/\alpha_4^2    \\
               0  & -\alpha_4\alpha_5+\alpha_5^4/\alpha_4^2 &  
\alpha_4\alpha_5- \alpha_5^4/\alpha_4^2   \\
 \end{pmatrix}.
\label{proto}
\end{eqnarray}
This neutrino matrix is a prototype 
which leads to  the tri-bimaximal mixing  
with the mass hierarchy $m_3\gg m_2\geq m_1$,
then we expect that realistic mass matrix is obtained near the condition
$y_1\alpha_5=y_2\alpha_4$. 

Let us discuss the detail of the mass matrix (\ref{mass}). 
After rotating $\theta_{23}=45^\circ$, we get
\begin{eqnarray}
 \frac{y_D^2v_u^2}{ \Lambda d}
\begin{pmatrix}y_1^2\alpha_5^2-y_2^2\alpha_4^2  & \sqrt2(-y_1y_2 
\alpha_5^2+y_2^2\alpha_4\alpha_5) & 0 \\ 
               \sqrt2(-y_1y_2 \alpha_5^2+y_2^2\alpha_4\alpha_5)   & y_1^2 
\alpha_4\alpha_5-y_1y_2 \alpha_4^2  &  0     \\
                 0   & 0  & y_1^2 \alpha_4\alpha_5+y_1y_2 
\alpha_4^2-2y_2^2\alpha_5^2   \\
 \end{pmatrix}, 
\end{eqnarray}
which  leads $\theta_{13}=0$ and
\begin{eqnarray}
\theta_{12}
=\frac12\arctan\frac{2\sqrt2 y_2\alpha_5}
{y_1\alpha_5+y_2\alpha_4-y_1 \alpha_4} \qquad (y_2\alpha_4\not 
=y_1\alpha_5).
\end{eqnarray}
Neutrino masses are given as
\begin{eqnarray}
m_1&=& \frac{y_D^2v_u^2}{ \Lambda d}
[y_1^2\alpha_5^2-y_2^2\alpha_4^2 -\sqrt2 (-y_1y_2 
\alpha_5^2+y_2^2\alpha_4\alpha_5)\tan\theta_{12}],
\nonumber\\
m_2&=& \frac{y_D^2v_u^2}{ \Lambda d}
[y_1^2 \alpha_4\alpha_5-y_1y_2 \alpha_4^2  + {\sqrt2}  (-y_1y_2 
\alpha_5^2+y_2^2\alpha_4\alpha_5) \tan\theta_{12}],
\nonumber\\
m_3&=&\ \frac{y_D^2v_u^2}{ \Lambda d}
[y_1^2 \alpha_4\alpha_5+y_1y_2 \alpha_4^2-2y_2^2\alpha_5^2],
\label{m3}
\end{eqnarray}
which are reconciled with the normal hierarchy of neutrino masses
in the case of $y_1\alpha_5\simeq y_2\alpha_4$.

Let us estimate magnitudes of $\alpha_i(i=4,5,6)$
by using Eq.(\ref{m3}). Suppose
$\tilde \alpha=\alpha_4\simeq \alpha_5=\alpha_6$.
If we take all Yukawa couplings to be order one,
Eq.(\ref{m3}) turns to be
$v_u^2=\Lambda \tilde\alpha m_3$ because of $d\sim \tilde\alpha^3$.
Putting $v_u\simeq 165$GeV ($\tan\beta=3$),
$m_3\simeq \sqrt{\Delta m_{\rm atm}^2}\simeq 0.05$eV,
and $\Lambda = 2.43\times 10^{18}$GeV, we obtain
$\tilde \alpha={\cal O}(10^{-4}-10^{-3})$.
Thus, values of $\alpha_i (i=4,5,6)$ are enough suppressed
to discuss perturbative series of higher mass operators.

\section{Higher order corrections}

Let us consider higher order contributions to mass matrices.
There are six expansion parameters $\alpha_i$, all of which
are expected to be small.

Since  products $3_1^{(1)}\times 3_2^{(2)}\times 2_1\times 1_2$
and  $3_1^{(1)}\times 3_2^{(2)}\times 2_1\times 2_1$
give the  $\Delta(54)$ invariant in the charged lepton sector,
the superpotential of next leading order is written as 
\begin{eqnarray}
\delta w_l
&=&
y_3^l
[(\omega  e^c l_e+\omega^2 \mu^c l_\mu+  \tau^c l_\tau )\chi_2 
+(e^c l_e+\omega^2 \mu^c l_\mu+\omega  \tau^c l_\tau )\chi_3]
\chi_1  h_d/\Lambda^2
\nonumber\\&&
+y_4^l 
( (e^c l_e+\omega^2 \mu^c l_\mu+\omega \tau^c l_\tau )\chi_2 ^2 - 
(\omega  e^c l_e+\omega^2  \mu^c l_\mu+ \tau^c l_\tau )\chi_3 ^2)
h_d/\Lambda^2. 
\label{LE}
\end{eqnarray}
For the right-handed Majorana neutrinos, 
 the  $\Delta(54)$ invariant product 
$3_1^{(2)}\times 3_1^{(2)}\times  3_1^{(2)}\times 2_1$ gives 
\begin{eqnarray}
\delta w_N
&=&
y_3
[(\omega N_e^c N_e^c\chi_4+\omega^2 N_\mu^c N_\mu^c\chi_5+
N_\tau^c N_\tau^c\chi_6)\chi_2
\nonumber\\&&
+( N_e^c N_e^c\chi_4+\omega^2 N_\mu^c N_\mu^c\chi_5+
\omega N_\tau^c N_\tau^c\chi_6)\chi_3 ]/\Lambda 
\nonumber\\&&
+y_4\left [\{\omega ( N_\mu^c N_\tau^c+ N_\tau^c N_\mu^c)\chi_4
+\omega^2(N_e^c N_\tau^c+ N_\tau^c N_e^c)\chi_5
+(N_e^c N_\mu^c+\bar N_\mu^c N_e^c)\chi_6\}\chi_2 \right .
\nonumber\\&&
+\{( N_\mu^c N_\tau^c+ N_\tau^c N_\mu^c)\chi_4
+\omega^2(N_e^c l_\tau^c+N_\tau^c N_e^c)\chi_5
\nonumber\\&&
+\omega ( N_e^c N_\mu+ N_\mu^c N_e^c)\chi_6\}\chi_3\left .\right ]
/\Lambda .
\label{LR}
\end{eqnarray}
The product $3_1^{(1)}\times 3_1^{(2)}\times 2_1$
gives a $\Delta(54)$ invariant in the Dirac neutrino sector as 
\begin{eqnarray}
\delta w_D
&=&
y^D_2 
[ (\omega  N_e^c l_e+\omega^2 N_ \mu^c l_\mu+ N_\tau^c  l_\tau )\chi_2
+ (N_e^c l_e+\omega^2 N_ \mu^c l_\mu+\omega  N_\tau^c l_\tau )\chi_3]
h_u/\Lambda.
\label{LD}
\end{eqnarray}
These correction terms of the superpotential in Eqs.
(\ref{LE}), (\ref{LR}), (\ref{LD}) give corrections of mass matrices 
\begin{eqnarray}
\delta M_l
 &= &
y_3^lv_d\alpha_1
\begin{pmatrix}\omega\alpha_2+\alpha_3  & 0 & 0 \\ 
           0  & \omega^2(\alpha_2+\alpha_3)  &  0  \\
                 0  & 0 & \alpha_2+\omega\alpha_3  \\
 \end{pmatrix} 
 \nonumber\\&&
+\ y_4^lv_d
\begin{pmatrix}\alpha_2^2-\omega\alpha_3^2  & 0 & 0 \\ 
           0  & \omega^2(\alpha_2^2-\alpha_3^2)  &  0  \\
                 0  & 0 & \omega\alpha_2^2-\alpha_3^2  \\
 \end{pmatrix}, 
\label{delE}
 \end{eqnarray}
for charged leptons,
\begin{eqnarray}
\delta M_N
&=&
y_3\Lambda
\begin{pmatrix}(\omega\alpha_2+\alpha_3)\alpha_4  & 0 & 0 \\ 
               0    & \omega^2(\alpha_2+\alpha_3)\alpha_5  &0    \\
                 0  & 0 &(\alpha_2+\omega\alpha_3)\alpha_6   \\
 \end{pmatrix} 
\nonumber\\&&
+\ y_4 \Lambda
\begin{pmatrix}0   & (\alpha_2+\omega\alpha_3)\alpha_{6} 
&\omega^2(\alpha_2+\alpha_3) \alpha_{5} \\ 
                   (\alpha_2+ \omega\alpha_3)\alpha_{6}  & 0   
&(\omega\alpha_2+\alpha_3)\alpha_{4}    \\
                   \omega^2(\alpha_2+ \alpha_3)\alpha_{5}  & 
(\omega\alpha_2+\alpha_3)\alpha_{4}  & 0    \\
 \end{pmatrix}, 
\label{delR}
\end{eqnarray}
for right-handed Majorana neutrinos, and
\begin{eqnarray}
\delta M_D
 = 
y^D_2v_u
\begin{pmatrix} \omega\alpha_2+ \alpha_3  & 0 & 0 \\ 
           0  &  \omega^2\alpha_2+\omega^2\alpha_3  &  0  \\
                 0  & 0 &  \alpha_2+\omega\alpha_3   \\
 \end{pmatrix},
\label{delD} 
 \end{eqnarray}
for Dirac neutrinos.
It is noticed that the corrections of the mass matrices
do not change the zero textures in the leading mass matrices
of Eqs. (\ref{ME}), (\ref{MR}), (\ref{MD}).

 Since the magnitudes  of $\alpha_i(i=1,2,3)$  are of 
${\cal   O}(10^{-2})$ as seen in
Eq.(\ref{alpha123}), mass matrix corrections
$\delta M_l$ and $\delta M_D$ in Eqs. (\ref{delE}) and (\ref{delD})
are suppressed enough.
On the other hand, the magnitudes of $\alpha_i(i=4,5,6)$ are
${\cal O}(10^{-4}-10^{-3})$ as discussed in the previous section.
Therefore, the correction $\delta M_N$ in Eq.(\ref{delR})
is also suppressed enough.
In conclusion, we can neglect the higher order contribution
in our numerical study of neutrino masses and mixing angles.


\section{Vacuum alignment}
We analyze the scalar potential to find out the vacuum alignment 
\footnote{Instead of analyzing the potential minimum, 
the vacuum alignment could be realized by imposing boundary conditions
of $\chi_i$ in extra dimensions \cite{Kobayashi:2008ih}
\cite{Seidl}.}. 
The scalar potential becomes rather simple in the $\Delta(54)$ symmetry.
Especially, the supersymmetry is important to see the vacuum 
alignment. 



The $\Delta(54)$ invariant superpotential is given as
\begin{eqnarray}
w &=& 
\mu _1\chi _1^2 + \mu _2\chi _2\chi _3 
\nonumber\\&&  
  + \eta _2(\chi _2^3+\chi _3^3) + \eta _3(\chi _4^3+\chi _5^3+\chi _6^3) + 
\eta _3^\prime \chi _4\chi _5\chi _6 
\nonumber\\&&  
  + \frac{\lambda _1}{\Lambda }\chi _1^4 + \frac{\lambda _2}{\Lambda }\chi 
_2^2\chi _3^2 
 + \frac{\lambda _3}{\Lambda }\chi _1^2\chi _2\chi _3 +\frac{\lambda 
_4}{\Lambda }\chi _1(\chi _2^3-\chi _3^3) 
\nonumber\\&&  
 + \frac{\lambda _6}{\Lambda }\left [\chi _2(\omega \chi _4^3+\omega ^2\chi 
_5^3+\chi _6^3)+\chi _3(\chi _4^3+\omega ^2\chi _5^3+\omega \chi 
_6^3)\right ]
\ ,
\end{eqnarray}
which leads to the   scalar potential 
\begin{eqnarray}
V &=& |2\mu _1\chi _1+4\frac{\lambda _1}{\Lambda }\chi _1^3+2\frac{\lambda 
_3}{\Lambda }\chi _1\chi _2\chi _3+\frac{\lambda _4}{\Lambda }(\chi 
_2^3-\chi _3^3)|^2
\nonumber\\
&& + |\mu _2\chi _3+3\eta _2\chi _2^2+2\frac{\lambda _2}{\Lambda }\chi 
_2\chi _3^2
+\frac{\lambda _3}{\Lambda }\chi _1^2\chi _3+3\frac{\lambda _4}{\Lambda 
}\chi _1\chi _2^2+\frac{\lambda _6}{\Lambda }(\omega \chi _4^3+\omega ^2\chi 
_5^3+\chi _6^3)|^2  
\nonumber\\
&& + |\mu _2\chi _2+3\eta _2\chi _3^2+2\frac{\lambda _2}{\Lambda }\chi 
_2^2\chi _3
+\frac{\lambda _3}{\Lambda }\chi _1^2\chi _2-3\frac{\lambda _4}{\Lambda 
}\chi _1\chi _3^2+\frac{\lambda _6}{\Lambda }(\chi _4^3+\omega ^2\chi 
_5^3+\omega \chi _6^3)|^2  
\nonumber\\
&& + |3\eta _3\chi _4^2+\eta _3^\prime \chi _5\chi _6+3\frac{\lambda 
_6}{\Lambda }(\omega \chi _2+\chi _3)\chi _4^2|^2  
+ |3\eta _3\chi _5^2+\eta _3^\prime \chi _4\chi _6+3\frac{\lambda 
_6}{\Lambda }\omega ^2(\chi _2+\chi _3)\chi _5^2|^2  
\nonumber\\
&& + |3\eta _3\chi _6^2+\eta _3^\prime \chi _4\chi _5+3\frac{\lambda 
_6}{\Lambda }(\chi _2+\omega \chi _3)\chi _6^2|^2 \ .
\end{eqnarray}
VEVs of $\chi_i$ must be much larger than the weak scale.
We assume that their VEVs are determined with neglecting 
supersymmetry breaking terms, i.e. $V_{\text{min}}=0$.
Then, the conditions of the potential minimum, $V_{\text{min}}=0$ are 
written as 
\begin{eqnarray}
2\mu _1\chi _1+4\frac{\lambda _1}{\Lambda }\chi _1^3+2\frac{\lambda 
_3}{\Lambda }\chi _1\chi _2\chi _3+\frac{\lambda _4}{\Lambda }(\chi _2^3-\chi 
_3^3) &=& 0 ,
\nonumber\\
\mu _2\chi _3+3\eta _2\chi _2^2+2\frac{\lambda _2}{\Lambda }\chi _2\chi 
_3^2
+\frac{\lambda _3}{\Lambda }\chi _1^2\chi _3+3\frac{\lambda _4}{\Lambda 
}\chi _1\chi _2^2+\frac{\lambda _6}{\Lambda }(\omega \chi _4^3+\omega ^2\chi 
_5^3+\chi _6^3) &=& 0 ,
\nonumber\\
 \mu _2\chi _2+3\eta _2\chi _3^2+2\frac{\lambda _2}{\Lambda }\chi _2^2\chi 
_3
+\frac{\lambda _3}{\Lambda }\chi _1^2\chi _2-3\frac{\lambda _4}{\Lambda 
}\chi _1\chi _3^2+\frac{\lambda _6}{\Lambda }(\chi _4^3+\omega ^2\chi 
_5^3+\omega \chi _6^3) &=& 0 ,
\nonumber\\
 3\eta _3\chi _4^2+\eta _3^\prime \chi _5\chi _6+3\frac{\lambda 
_6}{\Lambda }(\omega \chi _2+\chi _3)\chi _4^2 &=& 0 ,
\nonumber\\
 3\eta _3\chi _5^2+\eta _3^\prime \chi _4\chi _6+3\frac{\lambda 
_6}{\Lambda }\omega ^2(\chi _2+\chi _3)\chi _5^2 &=& 0 ,
\nonumber\\
 3\eta _3\chi _6^2+\eta _3^\prime \chi _4\chi _5+3\frac{\lambda 
_6}{\Lambda }(\chi _2+\omega \chi _3)\chi _6^2 &=& 0.
\end{eqnarray}
A solution of the last three equations is
\begin{eqnarray}
 \chi_4=\chi_5=\chi_6   \qquad  {\rm with} \qquad
 3\eta _3 + \eta _3^\prime = 0\  ,
\end{eqnarray}
where the higher dimensional operators proportional to $\lambda_6$
are neglected.
If we include the $\lambda_6$ terms,
the relation $\alpha_4=\alpha_5=\alpha_6$ is deviated in order of ${\cal O}(\alpha_i^2)$.
Therefore, we take randomly $\alpha_i(i=4,5,6)$ around
$\alpha_4=\alpha_5=\alpha_6$ in our numerical analysis.

\section{Numerical result}

We  show our numerical analysis of neutrino masses and mixing angles
in the normal mass hierarchy.
Neglecting  higher order corrections of mass matrices in section 3,
we obtain the allowed region of parameters and predictions of neutrino masses and mixing angles. Here, we neglect the renomarization effect of the 
neutrino mass matrix because we suppose the normal hierarchy of 
neutrino masses and take $\tan\beta = 3$.
\begin{figure}[tbh]
\begin{center}
\includegraphics[width=7 cm]{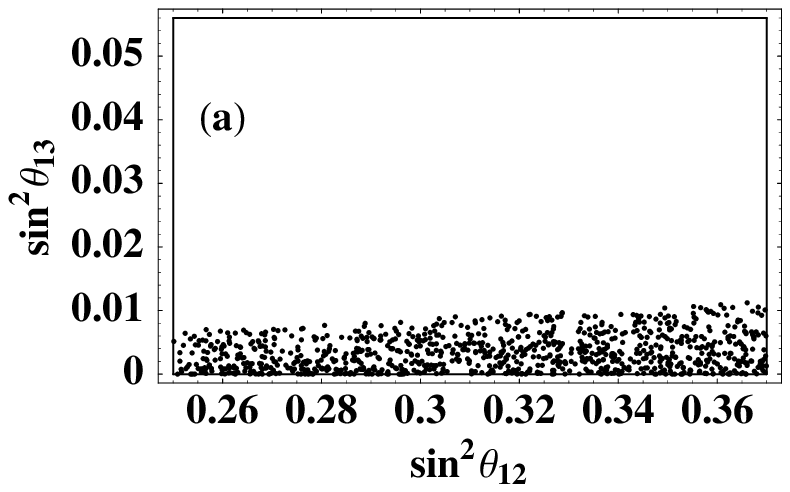}
\quad
\includegraphics[width=7 cm]{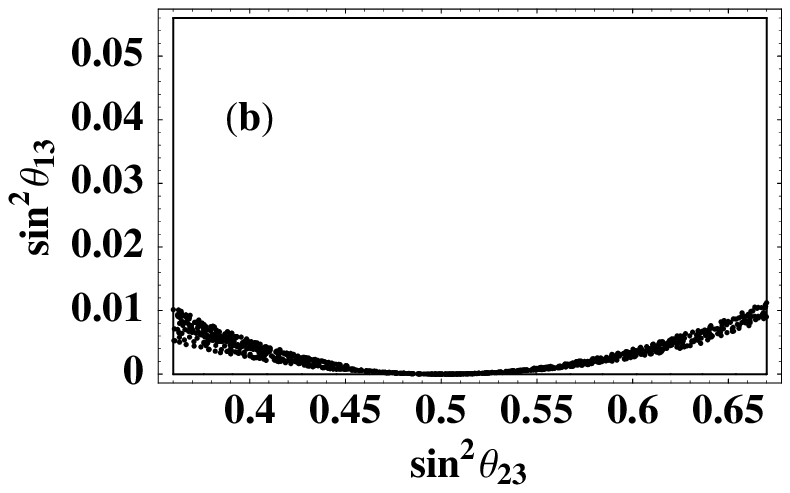}
\quad
\caption{Prediction of the upper bound of  $\sin^2\theta_{13}$
on (a) $\sin^2\theta_{12}-\sin^2\theta_{13}$ and
 (b) $\sin^2\theta_{23}-\sin^2\theta_{13}$ planes.}
\end{center}
\end{figure}
\begin{figure}[tb]
\begin{minipage}[]{0.4\linewidth} 
\includegraphics[width=7cm]{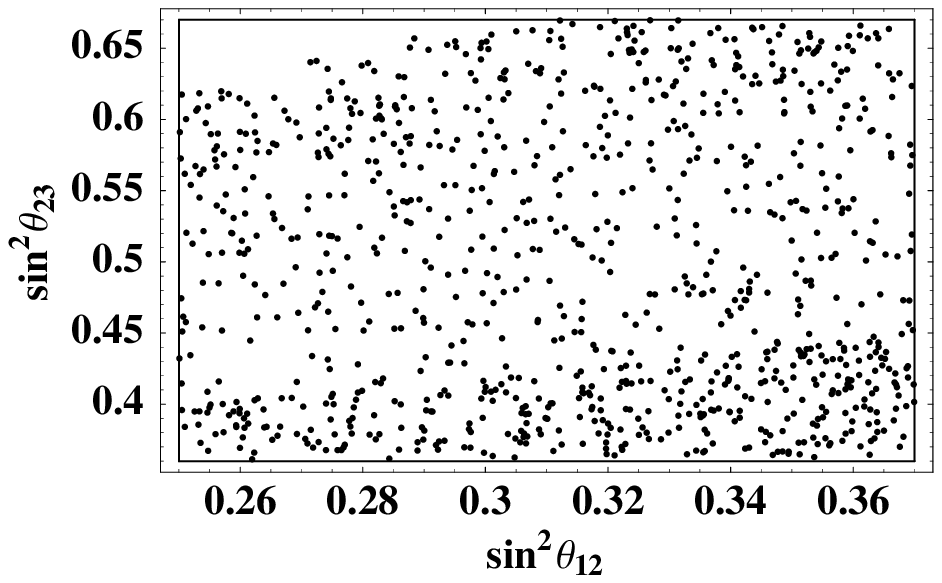}
\caption{Allowed region on  $\sin^2\theta_{12}-\sin^2\theta_{23}$ plane. }
\end{minipage}
\hspace{2cm}
\begin{minipage}[]{0.4\linewidth} 
\includegraphics[width=7 cm]{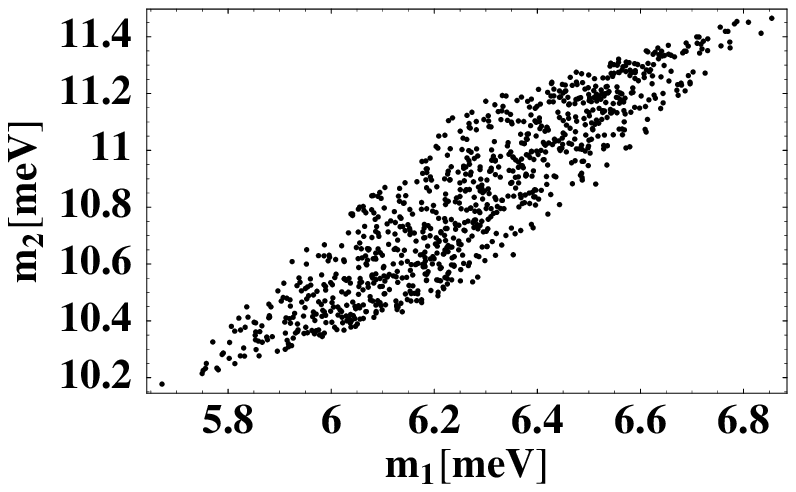}
\caption{The allowed mass region on the $m_1-m_2$ plane. }
\end{minipage}
\end{figure}

Input data of masses and mixing angles are taken in the  region of 
 3$\sigma$ of the experimental data \cite{Threeflavors}:
\begin{eqnarray}
&&\Delta m_{\rm atm}^2=(2.07\sim 2.75)\times 10^{-3} 
{\rm eV}^2 \ ,
\quad \Delta m_{\rm sol}^2= (7.05\sim 8.34) \times 10^{-5} {\rm eV}^2  \ , 
\nonumber \\
&& \sin^2 \theta_{\rm atm}=0.36\sim 0.67 \ ,
\quad  \sin^2 \theta_{\rm sol}=0.25 \sim 0.37  \ , \quad
 \sin^2 \theta_{\rm reactor} \leq 0.056\ ,
\end{eqnarray}
and   $\Lambda=2.43 \times 10^{18}$GeV is taken.
We fix $y_D=y_1=1$ as a convention, and vary $y_2/y_1$.
The change of $y_D$ and $y_1$  is absorbed into  the change of 
$\alpha_i(i=4,5,6)$.
If we take a smaller value of $y_1$, values of  $\alpha_i$ scale up.
On the other hand,
if we take a smaller value of $y_D$, the magnitude of $\alpha_i$ scale down.
As expected in the discussion of section 2,
the experimentally allowed values are reproduced
around $\alpha_4=\alpha_5=\alpha_6$.

 We can predict the deviation from the tri-bimaximal mixing.
The remarkable prediction is given in the magnitude of $\sin^2\theta_{13}$.
 In Figures 1 (a) and (b), we plot the allowed region of mixing angles
 in planes of $\sin^2\theta_{12}$-$\sin^2\theta_{13}$ and
$\sin^2\theta_{23}$-$\sin^2\theta_{13}$, respectively.
It is found that the upper bound of $\sin^2\theta_{13}$ is $0.01$.
It is also found the strong correlation between
$\sin^2\theta_{23}$ and $\sin^2\theta_{13}$.
Unless $\theta_{23}$ is deviated from the maximal mixing considerably,
$\theta_{13}$ remains to be tiny. This is a testable relation in this model.

The allowed region on the
$\sin^2\theta_{12}$-$\sin^2\theta_{23}$ plane is presented in Figure 2.
There is no correlation between $\sin^2\theta_{23}$ and $\sin^2\theta_{12}$
as well as between $\sin^2\theta_{13}$ and $\sin^2\theta_{12}$.

Let us discuss the first and second neutrino masses on the $m_1-m_2$ plane 
in Figure 3.
 We find   the lightest neutrino mass $m_1$ in the very narrow region 
of $m_1\simeq (6-7)\times 10^{-3}$eV in our model.

\begin{figure}[!tb]
\begin{center}
\includegraphics[width=7 cm]{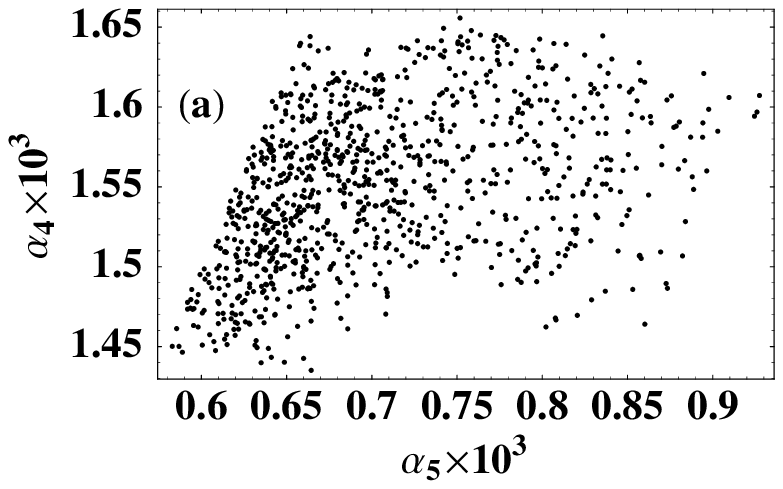}
\qquad
\includegraphics[width=7 cm]{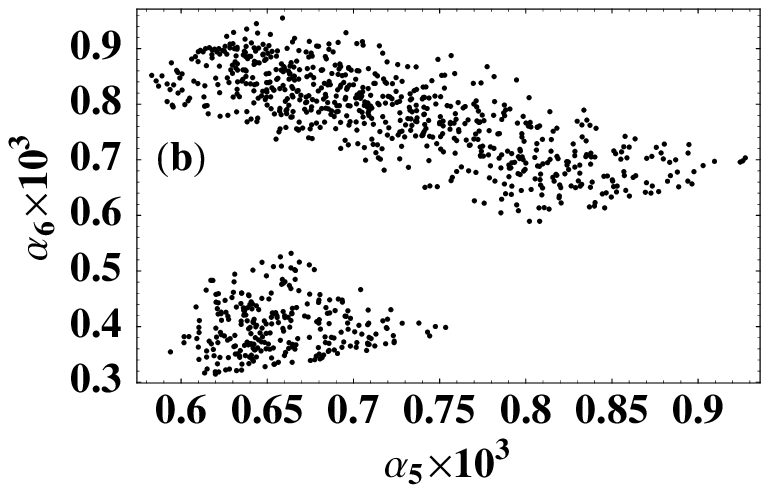}
\caption{Allowd regions on (a) $\alpha_5-\alpha_4$ and 
(b) $\alpha_5-\alpha_6$ planes. }
\end{center}
\end{figure}
\begin{wrapfigure}{r}{ 7.5cm}
\begin{center}
\includegraphics[width=7. cm]{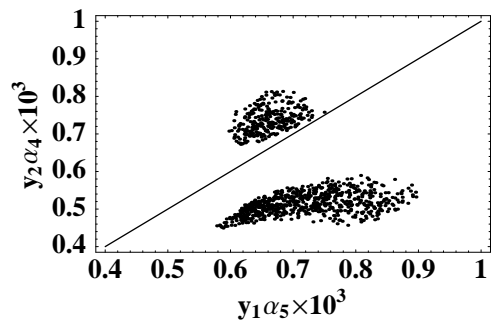}
\caption{The allowed region on  $y_1\alpha_5-y_2\alpha_4$ plane.
The solid line denote  $y_1\alpha_5=y_2\alpha_4$ one.}
\end{center}
\end{wrapfigure}

In Figure 4, we present allowed regions of parameters of
 $\alpha_4$, $\alpha_5$ and  $\alpha_6$, which give the
 neutrino masses and mixing angles consistent with the experimental data.
It is found  $\alpha_4\sim \alpha_5\sim \alpha_6 \sim{\cal O}(10^{-3})$,
which can be realized  in  the potential analysis of the previous section.
Since  the magnitude of $\alpha_i$ is found to be
 ${\cal O}(10^{-3})$ as expected in the section 2, 
 the neglect of the higher order corrections on the mass matrices
are guaranteed.

At last, we discuss about the relation of  $y_1\alpha_5\simeq y_2\alpha_4$,
which is expected in our analysis
 as discussed in Eq.(\ref{proto}) of section 2.
This relation is well  satisfied in our numerical result,
 which is shown on  $y_1\alpha_5- y_2\alpha_4$ plane in Figure 5.
By taking account of both results in  Figure 4(a) and Figure 5,
we have  found that the ratio $y_2/y_1$ is constrained  around $0.3- 0.5$.
Thus, Yukawa couplings $y_1$ and $y_2$ are of the same order.

\section{Summary and Discussion}

We have presented the  flavor model for the lepton mass matrices
by using the discrete symmetry $\Delta (54)$, which could be
originated from the string orbifold.
The left-handed leptons, the right-handed charged leptons
and the right-handed neutrinos are assigned by $3_1^{(1)}$,
$3_2^{(2)}$, and $3_1^{(2)}$, respectively.
We introduce  gauge singlets $\chi_1$, $(\chi_2, \chi_3)$ and
$(\chi_4, \chi_5, \chi_6)$, which are assigned to be 
$1_2$, $2_1$, and $3_1^{(2)}$ of the $\Delta(54)$ representations,
respectively.
The $\Delta(54)$ flavor symmetry can appear 
in heterotic string models on factorizable orbifolds 
including the $T^2/Z_3$ orbifold \cite{Kobayashi:2006wq}.
In these string models only singlets and 
triplets appear as fundamental modes, 
but doublets do not appear as fundamental modes.
The doublet plays an role in our model, 
and such doublet could appear, e.g. 
as composite modes of triplets.

As discussed in Eqs.(\ref{tri-bi1})-(\ref{tri-bi2}), the tri-bimaximal mixing is not realized for arbitrary neutrino masses  in our model.
Parameters are adapted to get neutrino masses  consistent 
with  observed values of $\Delta m^2_{\rm atm}$ and $\Delta m_{\rm sol}^2$.
Then, the deviation from  the tri-bimaximal mixing is estimated.
Therefore, our  approach does not predict the tri-bimaximal mixing,
but constrain the neutrino mass matrix by putting $\theta_{23}\simeq \pi/4$
by hand.

It is useful to give a following comment as to $\Delta(27)$ flavor symmetry.
Our mass matrix gives  the same result 
in the   $\Delta(27)$ flavor symmetry  \cite{Delta(27)-Ma(2006)},
where the type II seesaw is used.
Our  neutrino   mass matrix  is given as  $M_\nu\propto  M_N^{-1}$, where
 $M_N$ is the just  same  as  the neutrino mass matrix $M_\nu$
in the $\Delta(27)$ flavor symmetry \cite{Delta(27)-Ma(2006)}.
Therefore, if the type I seesaw is used in the $\Delta(27)$ flavor symmetry,
the same neutrino mass matrix  can be obtained.



The model  reproduces the almost tri-bimaximal mixing in 
the parameter region around two vanishing neutrino masses.
We have  predicted  the deviation from the tri-bimaximal mixing
 by input of the experimental data of $\Delta m^2_{\rm atm}$
 and  $\Delta m^2_{\rm sol}$ 
in the case of normal hierarchy of neutrino masses.
We have found that the upper bound of $\sin^2\theta_{13}$ is $0.01$.
There is the strong correlation between
$\sin^2\theta_{23}$ and $\sin^2\theta_{13}$.
Unless $\theta_{23}$ is deviated from the maximal mixing considerably,
$\theta_{13}$ remains to be tiny.
Therefore, the  model is testable in the future neutrino experiments.

\vspace{1cm}
\noindent
{\bf Acknowledgement}

T.~K.\/ is supported in part by the
Grant-in-Aid for Scientific Research, No.~20540266,
and 
the Grant-in-Aid for the Global COE Program 
"The Next Generation of Physics, Spun from Universality 
and Emergence" from the Ministry of Education, Culture,
Sports, Science and Technology of Japan.
The work of M.T. has been  supported by the
Grant-in-Aid for Science Research
of the Ministry of Education, Science, and Culture of Japan
No. 17540243.

\newpage
\appendix
\section{Appendix}
\subsection{Character table of $\Delta(54)$}

Group-theoretical aspects of $\Delta(54)$ can be found 
in ref.\cite{delta54}, in which  $\Delta(6n^2)$ is investigated.
 $\Delta(54)$ is a discrete subgroup of 
$\mathrm{SU}(3)$, i.e.\ the group $\Delta(6n^2)$ (with $n=3$) and 
it has order 
$54$. 
The generators of $\Delta(54)$ are given by the set 
\begin{equation}
a= \left(
 \begin{array}{ccc}
 0 & 1 & 0 \\
 0 & 0 & 1 \\
 1 & 0 & 0
 \end{array}
 \right),  \
b= \left(
 \begin{array}{ccc}
 0 & 0 & 1 \\
 0 & 1 & 0 \\
 1 & 0 & 0
 \end{array}
 \right),  \
c= \left(
 \begin{array}{ccc}
 \omega & 0 & 0 \\
 0 &  \omega^2 & 0 \\
 0 & 0 & 1
 \end{array}
 \right), \
c'=  \left(
 \begin{array}{ccc}
 \omega^2 & 0 & 0 \\
 0 &  \omega & 0 \\
 0 & 0 & 1
 \end{array}
 \right).
 \end{equation}
It has four three-dimensional irreducible representations 
$\mathbf{3}_1^{(1)},\,
\mathbf{{3}}_1^{(2)}, \, \mathbf{3}_2^{(1)},\, \mathbf{{3'}}_2^{(2)}\,$, 
four two-dimensional
ones $\mathbf{2}_{1},\, \mathbf{2}_{2}, \, \mathbf{2}_{3}, \, 
\mathbf{2}_{4}$,
and two one-dimensional ones $\mathbf{1}_{1}, \, \mathbf{1}_{2}$. 
Generators of three-dimensional representations are mainly divided into two 
types. 
For $\mathbf{{3}}_1^{(1)}, ~\mathbf{{3}}_1^{(2)}$, generators are $a$, $b$, 
$c$, and 
for $\mathbf{{3}}_2^{(1)}, ~\mathbf{{3}}_2^{(2)}$, generators are $a$, $b$, 
$c'$.
Their
character table  are presented  in Table \ref{character}. 

\vskip 1 cm
\begin{table}[!h]
\centerline{
\begin{tabular}{|c|rrrrrrrrrr|}
\hline
irrep & 1a  & 6a  & 6b  & 3a  & 3b  & 3c  & 2a  & 3d  & 3e  & 3f \\
      & \scriptsize{(1)} & \scriptsize{(9)} & \scriptsize{(9)} & 
\scriptsize{(6)} & \scriptsize{(6)} & \scriptsize{(6)} & \scriptsize{(9)} & 
\scriptsize{(6)} & \scriptsize{(1)} & \scriptsize{(1)} \\
\hline
$\mathbf{1}_{1}$ & 1 & 1 & 1 & 1 & 1 & 1 & 1 & 1 & 1 & 1 \\
$\mathbf{1}_{2}$ & 1 & -1 & -1 & 1 & 1 & 1 & -1 & 1 & 1 & 1 \\
$\mathbf{2}_{1}$ & 2 & 0 & 0 & 2 & -1 & -1 & 0 & -1 & 2 & 2 \\
$\mathbf{2}_{2}$ & 2 & 0 & 0 & -1 & -1 & -1 & 0 & 2 & 2 & 2 \\
$\mathbf{2}_{3}$ & 2 & 0 & 0 & -1 & -1 & 2 & 0 & -1 & 2 & 2 \\
$\mathbf{2}_{4}$ & 2 & 0 & 0 & -1 & 2 & -1 & 0 & -1 & 2 & 2 \\
$\mathbf{3_2^{(1)}}$ & 3 & $-\bar{\omega}$ & $-\omega$ & 0 & 0 & 0 & -1 & 0 
& $3\bar{\omega}$ & $3\omega$ \\
$\mathbf{3_2^{(2)}}$ & 3 & $-\omega$ & $-\bar{\omega}$ & 0 & 0 & 0 & -1 & 0 
& $3\omega$ & $3\bar{\omega}$ \\
$\mathbf{3_1^{(2)}}$ & 3 & $\omega$ & $\bar{\omega}$ & 0 & 0 & 0 & 1 & 0 & 
$3\omega$ & $3\bar{\omega}$ \\
$\mathbf{3_1^{(1)}}$ & 3 & $\bar{\omega}$ & $\omega$ & 0 & 0 & 0 & 1 & 0 & 
$3\bar{\omega}$ & $3\omega$ \\
\hline
\end{tabular}
}
\caption{Character table of the group $\Delta(54)$. }
\label{character}
\end{table}

\newpage

\subsection{Kronecker products}
We display Kronecker products and calculation of Clebsch Gordan coefficients.
The Kronecker products can be calculate from the character table 
in the previous subsection.             
\begin{equation}
\begin{split}
&1_i\times 1_i = 1_1\ (i=1,2), \quad 1_1\times 1_2 = 1_2\times 1_1 = 1_2, 
\\
&1_i\times 2_r = 2_r, \quad 1_i \times 3_j^{(l)} = 3_{((i+j)\bmod 
2)+1}^{(l)}\ (j,l=1,2), \\
&2_r\times 2_r = 1_1 + 1_2 + 2_r\ (r=1,2,3,4), \\
&2_a\times 2_b = 2_c + 2_d\ (a,b,c,d = 1,2,3,4,\ \text{different each other}), \\
&2_r\times 3_j^{(l)} = 3_1^{(l)} + 3_2^{(l)}, \\
&3_j^{(l)}\times 3_j^{(l)} = 3_1^{(l^\prime )} + 3_1^{(l^\prime )} + 
3_2^{(l^\prime )}\ (l^\prime =1,2,\ l\not = l' ), \\
&3_j^{(l)}\times 3_{j^\prime }^{(l)} = 3_2^{(l^\prime )} + 3_2^{(l^\prime 
)} + 3_1^{(l^\prime )}\ (j^\prime =1,2, \ j\not =j' ), \\
&3_j^{(l)}\times 3_j^{(l^\prime )} = 1_1 + 2_1 + 2_2 + 2_3 + 2_4, \\
&3_j^{(l)}\times 3_{j^\prime }^{(l^\prime )} = 1_2 + 2_1 + 2_2 + 2_3 + 2_4.
\end{split}
\end{equation}

\subsection{Multiplication of $\Delta(54)$}
We present the  relevant  multiplication rules 
of $\Delta (54)$. 
The multiplication rules of two dimensional representation
are given as
\begin{align*}
(x_1,x_2)_{2_r}\times (y_1,y_2)_{2_r} &= (x_1y_2+x_2y_1)_{1_1} + 
(x_1y_2-x_2y_1)_{1_2} + (x_2y_2,x_1y_1)_{2_r} \ (r=1,2,3,4) \\
(x_1,x_2)_{2_1}\times (y_1,y_2)_{2_2} &= (x_2y_2,x_1y_1)_{2_3} + 
(x_2y_1,x_1y_2)_{2_4} \\
(x_1,x_2)_{2_1}\times (y_1,y_2)_{2_3} &= (x_2y_2,x_1y_1)_{2_2} + 
(x_2y_1,x_1y_2)_{2_4}, \\
(x_1,x_2)_{2_1}\times (y_1,y_2)_{2_4} &= (x_1y_2,x_2y_1)_{2_2} + 
(x_1y_1,x_2y_2)_{2_3}, \\
(x_1,x_2)_{2_2}\times (y_1,y_2)_{2_3} &= (x_2y_2,x_1y_1)_{2_1} + 
(x_1y_2,x_2y_1)_{2_4},\\
(x_1,x_2)_{2_2}\times (y_1,y_2)_{2_4} &= (x_1y_1,x_2y_2)_{2_1} + 
(x_1y_2,x_2y_1)_{2_3}, \\
(x_1,x_2)_{2_3}\times (y_1,y_2)_{2_4} &= (x_1y_2,x_2y_1)_{2_1} + 
(x_1y_1,x_2y_2)_{2_2}.
\end{align*}
The multiplication rules of three dimensional representation is given as
\begin{align*}
(x_1,x_2,x_3)_{3_1^{(1)}}\times (y_1,y_2,y_3)_{3_1^{(1)}} &= 
(x_1y_1,x_2y_2,x_3y_3)_{3_1^{(2)}} \\
&\ \ \ \ + (x_2y_3+x_3y_2,x_3y_1+x_1y_3,x_1y_2+x_2y_1)_{3_1^{(2)}} \\
&\ \ \ \ + (x_2y_3-x_3y_2,x_3y_1-x_1y_3,x_1y_2-x_2y_1)_{3_2^{(2)}}, \\
(x_1,x_2,x_3)_{3_1^{(2)}}\times (y_1,y_2,y_3)_{3_1^{(2)}} &= 
(x_1y_1,x_2y_2,x_3y_3)_{3_1^{(1)}} \\
&\ \ \ \ + (x_2y_3+x_3y_2,x_3y_1+x_1y_3,x_1y_2+x_2y_1)_{3_1^{(1)}} \\
&\ \ \ \ + (x_2y_3-x_3y_2,x_3y_1-x_1y_3,x_1y_2-x_2y_1)_{3_2^{(1)}}, \\
(x_1,x_2,x_3)_{3_1^{(1)}}\times (y_1,y_2,y_3)_{3_1^{(2)}} &= 
(x_1y_1+x_2y_2+x_3y_3)_{1_1} \\
&\ \ \ \ + (x_1y_1+\omega ^2x_2y_2+\omega x_3y_3,\omega x_1y_1+\omega 
^2x_2y_2+x_3y_3)_{2_1} \\
&\ \ \ \ + (x_1y_2+\omega ^2x_2y_3+\omega x_3y_1,\omega x_1y_3+\omega 
^2x_2y_1+x_3y_2)_{2_2} \\
&\ \ \ \ + (x_1y_3+\omega ^2x_2y_1+\omega x_3y_2,\omega x_1y_2+\omega 
^2x_2y_3+x_3y_1)_{2_3} \\
&\ \ \ \ + (x_1y_3+x_2y_1+x_3y_2,x_1y_2+x_2y_3+x_3y_1)_{2_4}, \\
(x_1,x_2,x_3)_{3_1^{(2)}}\times (y_1,y_2,y_3)_{3_1^{(1)}} &= 
(x_1y_1+x_2y_2+x_3y_3)_{1_1} \\
&\ \ \ \ + (x_1y_1+\omega ^2x_2y_2+\omega x_3y_3,\omega x_1y_1+\omega 
^2x_2y_2+x_3y_3)_{2_1} \\
&\ \ \ \ + (x_1y_3+\omega ^2x_2y_1+\omega x_3y_2,\omega x_1y_2+\omega 
^2x_2y_3+x_3y_1)_{2_2} \\
&\ \ \ \ + (x_1y_2+\omega ^2x_2y_3+\omega x_3y_1,\omega x_1y_3+\omega 
^2x_2y_1+x_3y_2)_{2_3} \\
&\ \ \ \ + (x_1y_2+x_2y_3+x_3y_1,x_1y_3+x_2y_1+x_3y_2)_{2_4}, \\
(x_1,x_2,x_3)_{3_2^{(1)}}\times (y_1,y_2,y_3)_{3_2^{(1)}} &= 
(x_1y_1,x_2y_2,x_3y_3)_{3_1^{(2)}} \\
&\ \ \ \ + (x_2y_3+x_3y_2,x_3y_1+x_1y_3,x_1y_2+x_2y_1)_{3_1^{(2)}} \\
&\ \ \ \ + (x_2y_3-x_3y_2,x_3y_1-x_1y_3,x_1y_2-x_2y_1)_{3_2^{(2)}}, \\
(x_1,x_2,x_3)_{3_2^{(2)}}\times (y_1,y_2,y_3)_{3_2^{(2)}} &= 
(x_1y_1,x_2y_2,x_3y_3)_{3_1^{(1)}} \\
&\ \ \ \ + (x_2y_3+x_3y_2,x_3y_1+x_1y_3,x_1y_2+x_2y_1)_{3_1^{(1)}} \\
&\ \ \ \ + (x_2y_3-x_3y_2,x_3y_1-x_1y_3,x_1y_2-x_2y_1)_{3_2^{(1)}}, \\
(x_1,x_2,x_3)_{3_1^{(1)}}\times (y_1,y_2,y_3)_{3_2^{(2)}} &= 
(x_1y_1+x_2y_2+x_3y_3)_{1_2} \\
&\ \ \ \ + (x_1y_1+\omega ^2x_2y_2+\omega x_3y_3,-\omega x_1y_1-\omega 
^2x_2y_2-x_3y_3)_{2_1} \\
&\ \ \ \ + (x_1y_2+\omega ^2x_2y_3+\omega x_3y_1,-\omega x_1y_3-\omega 
^2x_2y_1-x_3y_2)_{2_2} \\
&\ \ \ \ + (x_1y_3+\omega ^2x_2y_1+\omega x_3y_2,-\omega x_1y_2-\omega 
^2x_2y_3-x_3y_1)_{2_3} \\
&\ \ \ \ + (x_1y_3+x_2y_1+x_3y_2,-x_1y_2-x_2y_3-x_3y_1)_{2_4},\\
(x_1,x_2,x_3)_{3_1^{(2)}}\times (y_1,y_2,y_3)_{3_2^{(1)}} &= 
(x_1y_1+x_2y_2+x_3y_3)_{1_2} \\
&\ \ \ \ + (x_1y_1+\omega ^2x_2y_2+\omega x_3y_3,-\omega x_1y_1-\omega 
^2x_2y_2-x_3y_3)_{2_1} \\
&\ \ \ \ + (x_1y_3+\omega ^2x_2y_1+\omega x_3y_2,-\omega x_1y_2-\omega 
^2x_2y_3-x_3y_1)_{2_2} \\
&\ \ \ \ + (x_1y_2+\omega ^2x_2y_3+\omega x_3y_1,-\omega x_1y_3-\omega 
^2x_2y_1-x_3y_2)_{2_3} \\
&\ \ \ \ + (x_1y_2+x_2y_3+x_3y_1,-x_1y_3-x_2y_1-x_3y_2)_{2_4}.
\end{align*}

\end{document}